\documentclass[8.5pt,twoside,twocolumn]{article}
\usepackage[super,sort&compress,comma]{natbib}
\usepackage{mhchem}
\usepackage{times,mathptmx}
\usepackage{sectsty}
\usepackage{balance}
\usepackage{multirow}
\usepackage{epsfig}
\usepackage{lastpage}
\usepackage[format=plain,justification=raggedright,singlelinecheck=false,font=small,labelfont=bf,labelsep=space]{caption}

\begin{document}

\thispagestyle{plain}
%\fancypagestyle{plain}{
%\fancyhead[L]{\includegraphics[height=8pt]{headers/LH}}
%\fancyhead[C]{\hspace{-1cm}\includegraphics[height=20pt]{headers/CH}}
%\fancyhead[R]{\includegraphics[height=10pt]{headers/RH}\vspace{-0.2cm}}
%\renewcommand{\headrulewidth}{1pt}}
%\renewcommand{\thefootnote}{\fnsymbol{footnote}}
%\renewcommand\footnoterule{\vspace*{1pt}%
%\hrule width 3.4in height 0.4pt \vspace*{5pt}}
%\setcounter{secnumdepth}{5}

\makeatletter
\def\subsubsection{\@startsection{subsubsection}{3}{10pt}{-1.25ex plus -1ex minus -.1ex}{0ex plus 0ex}{\normalsize\bf}}
\def\paragraph{\@startsection{paragraph}{4}{10pt}{-1.25ex plus -1ex minus -.1ex}{0ex plus 0ex}{\normalsize\textit}}
\renewcommand\@biblabel[1]{#1}
\renewcommand\@makefntext[1]%
{\noindent\makebox[0pt][r]{\@thefnmark\,}#1}
\makeatother
\renewcommand{\figurename}{\small{Fig.}~}
\sectionfont{\large}
\subsectionfont{\normalsize}

%\fancyfoot{}
%\fancyfoot[LO,RE]{\vspace{-7pt}\includegraphics[height=9pt]{headers/LF}}
%\fancyfoot[CO]{\vspace{-7.2pt}\hspace{12.2cm}\includegraphics{headers/RF}}
%\fancyfoot[CE]{\vspace{-7.5pt}\hspace{-13.5cm}\includegraphics{headers/RF}}
%\fancyfoot[RO]{\footnotesize{\sffamily{1--\pageref{LastPage} ~\textbar  \hspace{2pt}\thepage}}}
%\fancyfoot[LE]{\footnotesize{\sffamily{\thepage~\textbar\hspace{3.45cm} 1--\pageref{LastPage}}}}
%\fancyhead{}
%\renewcommand{\headrulewidth}{1pt}
%\renewcommand{\footrulewidth}{1pt}
%\setlength{\arrayrulewidth}{1pt}
%\setlength{\columnsep}{6.5mm}
%\setlength\bibsep{1pt}

\twocolumn[
  \begin{@twocolumnfalse}
\noindent\LARGE{\textbf{Order and disorder around Cr$^{3+}$ in chromium doped persistent luminescence AB$_2$O$_4$ spinels}}
\vspace{0.6cm}

\noindent\large{\textbf{Neelima Basavaraju,\textit{$^{a}$} Kaustubh R. Priolkar,$^{\ast}$\textit{$^{a}$} Didier Gourier,\textit{$^{b}$} Aur\'{e}lie Bessi\`{e}re,\textit{$^{b}$} and Bruno Viana,\textit{$^{b}$}}}\vspace{0.5cm}
%Please note that \ast indicates the corresponding author(s) but no footnote text is required.

\noindent\textit{\small{\textbf{Received Xth XXXXXXXXXX 20XX, Accepted Xth XXXXXXXXX 20XX\newline
First published on the web Xth XXXXXXXXXX 200X}}}

\noindent \textbf{\small{DOI: 10.1039/b000000x}}
\vspace{0.6cm}
%Please do not change this text.

\noindent \normalsize{X-ray absorption near edge structure (XANES) spectroscopy technique is used to better understand the charging and decharging processes of the persistent luminescence in the Cr$^{3+}$ doped AB$_2$O$_4$ spinels (A = Zn, Mg and B = Ga and Al) with low photon energy excitation by visible light. Cr K edge XANES spectra have been simulated for different near neighbour environments around the Cr$^{3+}$ recombination centres and compared with the experimental curve. In Cr$^{3+}$:ZnGa$_2$O$_4$ compound, the Cr$^{3+}$ local structure corresponds mostly to that of a normal spinel ($\sim$70\%), while the rest comprises of distorted octahedral environment arising from cationic site inversion and a contribution from chromium clustering. This local structure is strongly different in Cr$^{3+}$:MgGa$_2$O$_4$ and Cr$^{3+}$:ZnAl$_2$O$_4$ where, for both cases, chromium clustering represents the main contribution. The strong correlation between the intensity of persistent luminescence and \% of Cr in clusters leads us to infer that presence of Cr clusters is responsible for decrease of the intensity of the visible light induced persistent luminescence in the Cr$^{3+}$ doped AB$_2$O$_4$ spinels.}
\vspace{0.5cm}
 \end{@twocolumnfalse}
  ]

%\section{This is the section heading style}
%Footnotes
\footnotetext{\dag~Electronic Supplementary Information (ESI) available: [details of any supplementary information available should be included here]. See DOI: 10.1039/b000000x/}

%Please use \dag to cite the ESI in the main text of the article.
%If you article does not have ESI please remove the the \dag symbol from the title and the above footnotetext.

\footnotetext{\textit{$^{a}$Department of Physics, Goa University, Taleigao plateau, Goa 403206, India. Tel: 0832 651 9084; E-mail: krp@unigoa.ac.in}}
\footnotetext{\textit{$^{b}$PSL Research University, Chimie ParisTech - CNRS, Institut de Recherche de Chimie Paris, 75005, Paris, France.}}

\section{Introduction}

Near infrared persistent luminescence, hereafter referred to as NIRPL, has been the object of utmost interest since the demonstration that NIRPL nanoparticles can be used for\textit{ in vivo} optical imaging.\cite{1} In the course of finding new materials with improved optical properties, it was found that Cr$^{3+}$ doped zinc gallate and gallogermanate spinel family is particularly interesting as these materials give enhanced NIRPL intensity and duration.\cite{2,3,4,5} In addition, the interest of these materials for application of \textit{in vivo} imaging is reinforced by the fact that NIRPL can be excited not only by UV light, but also by visible light\cite{3}, which allowed direct excitation of ZnGa$_2$O$_4$:Cr$^{3+}$ nanoparticles through animal body for \textit{in vivo} tumour imaging\cite{5}.  Recently, Zhuang et al. proposed very clearly a mechanism for NIRPL in Cr doped zinc gallate, showing the different pathways by which an optically excited Cr$^{3+}$ transfers an electron to a shallow trap in the band gap.\cite{6} However it appears that the actual mechanism is probably more complex. First, Cr$^{3+}$ seems to conserve its oxidation state during excitation of NIRPL.\cite{7} Second, photoluminescence excitation and emission NIRPL spectra show that NIRPL mechanism is due to a special type of Cr$^{3+}$ with neighbouring defects, mostly antisite defects.\cite{7} The complexity of the environment of Cr$^{3+}$ in ZnGa$_2$O$_4$ has been shown by EPR spectroscopy.\cite{8} All these results show that NIRPL mechanism in chromium doped zinc gallate is determined by the local environment of chromium.

Synchrotron based X-ray absorption fine structure (XAFS) spectroscopy techniques are being widely used now to study the local structure around a selected element and thus to determine the structure of the given material.\cite{9,10,11,12,13,14} Tunable, coherent, polarized radiation along with a high photon flux obtained from a synchrotron makes it favourable over conventional spectroscopic methods.\cite{15,16} Owing to the applicability of XAFS to crystalline as well as amorphous materials, the technique can be employed to characterize a wide variety of materials such as optical materials,\cite{9,11,17} magnetic materials,\cite{13} ancient and historical materials,\cite{15} et cetera. Based on interference of forward going wave of photoelectron (which is emitted due to atomic absorption of incident photon) and back scattered wave of photoelectron from neighbouring ions, XAFS provides information about immediate surroundings of the central absorbing ion.\cite{18} Therefore it can be used to probe structural defects around a metal ion. A XAFS spectrum is divided into two parts – X-ray absorption near edge structure (XANES) and extended X-ray absorption fine structure (EXAFS). The spectrum up to about 30 eV from the edge energy where the photoelectron scattering is strong is the XANES part. 

XAFS has been used by many groups in recent times to study the local structure around luminescent ions in order to understand the mechanism of persistent luminescence. Extensive studies done on Eu$^{2+}$ doped Sr$_2$MgSi$_2$O$_7$ in 2009 unanimously showed that electron is excited to conduction band from Eu$^{2+}$ ion which then acts as a hole trap.\cite{19,20,21} In case of rare earth trivalent ion (R$^{3+}$) co-doped Sr$_2$MgSi$_2$O$_7$:Eu$^{2+}$ it was proposed that R$^{3+}$ ions with energy levels just below conduction band were acting as electron traps, later releasing the electron to recombine with the hole.\cite{19} Similar mechanism was extended for CaAl$_2$O$_4$:Eu$^{2+}$,R$^{3+}$ compound, proposing the conversion of Eu$^{2+}$ to Eu$^{3+}$ with the electron transfer.\cite{22} However this valence state change was disproved by the same group when they observed no signature of Eu$^{3+}$ oxidation state in the XANES spectra.\cite{23} Korthout et. al. carried out XANES studies on SrAl$_2$O$_4$:Eu,Dy to show the valence state change of Eu$^{2+}$ to Eu$^{3+}$ during the charging process, but no valence change in Dy was seen.\cite{24} XAFS measurements at Ca K edge and Mn K edge in CaMgSi$_2$O$_6$:Mn helped to establish the formation of oxygen vacancies around Ca when the compounds were prepared in reducing atmosphere, which act as electron traps while the luminescent centre Mn was the hole trap.\cite{9,11} Rodrigues et. al. showed from XAFS investigations that there is no oxidation state change of Tb$^{3+}$ in CdSiO$_3$:Tb$^{3+}$ during the excitation process.\cite{25}

In a previous paper, we had presented EXAFS results along with optical and electron paramagnetic resonance (EPR) studies on Cr$^{3+}$ doped AB$_2$O$_4$ spinels (A=Zn, Mg and B=Ga, Al).\cite{17} In there, the crucial role of inversion defects in necessitating the persistent luminescence in spinels specially with visible light excitation was illustrated in detail. A strong correlation between Cr-O bond distances and visible light induced NIRPL was shown.\cite{17} This was interpreted to be due to increased hybridization between Cr 3d t$_{2g}$ orbitals and ligand 2p orbitals due to presence of antisite defects in the near neighbourhood of Cr$^{3+}$ ions. The $\pi$ bonding between Cr and O resulted in decreased crystal field (CF) splitting of the Cr 3d band in spite of decrease in Cr-O bond distance in these Cr$^{3+}$ doped AB$_2$O$_4$ spinels.  Since XANES is very sensitive to CF splitting, nature of bonding and near neighbour environment, we present here our investigations on Cr K edge XANES spectra in Cr$^{3+}$ doped AB$_2$O$_4$ spinels  (A = Zn, Mg and B = Ga and Al). \textit{Ab initio} calculations of Cr XANES spectra for each compound have been performed for different Cr$^{3+}$ environments with the introduction of defects possibly identified by optical spectroscopy and EPR\cite{7,8}, in order to identify the role of these defects in persistent luminescence property of these spinels when low photon energy (visible light) is used during the charging process.

\begin{figure}[h]
\centering
\includegraphics[width=\columnwidth]{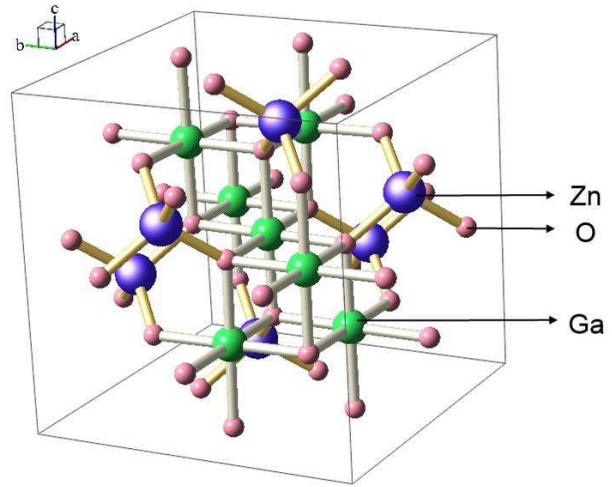}
\caption{Spinel structure of ZnGa$_2$O$_4$.}
\label{fig1}
\end{figure}

The spinels being discussed here (ZnGa$_2$O$_4$, MgGa$_2$O$_4$ and ZnAl$_2$O$_4$) are semiconductors with a wide energy gap which allows the possibility to create defect levels within bandgap by doping with transition metal ions. They belong to cubic space group \textit{Fd3m} with a reported lattice parameter a = 8.334 \AA ~for ZnGa$_2$O$_4$\cite{26}, a = 8.2891 \AA ~for MgGa$_2$O$_4$\cite{27} and a = 8.086 \AA ~for ZnAl$_2$O$_4$\cite{28}. ZnGa$_2$O$_4$ compound possesses a normal spinel structure with Zn$^{2+}$ ions in tetrahedral coordination and Ga$^{3+}$ ions in octahedral coordination (Figure \ref{fig1}),\cite{29} with a small inversion in site occupancies of Zn and Ga.\cite{30,31,32} It exhibits longest NIRPL signal as compared to the other two spinel hosts in the case of low photon energy excitation by visible light.\cite{17} MgGa$_2$O$_4$ is reported to be a near inverse spinel with about 44\% octahedral site inversion.\cite{27,33} ZnAl$_2$O$_4$ is known to crystallize in a perfect normal spinel structure with less than 1\% cationic disorder.\cite{34} While these three spinels give strong NIRPL with short excitation wavelength, ZnAl$_2$O$_4$ does not give any detectable NIRPL when excited by visible light.\cite{17}

\section{Experimental}

The samples under study were prepared via solid state route with their respective metal oxides ZnO (Sigma Aldrich 99.99\% pure), Ga$_2$O$_3$ (Sigma Aldrich 99.99\% pure), MgO (Sigma Aldrich 99.995\% pure), Al$_2$O$_3$ (SRL 99.75\% pure) and CrO$_3$ (SRL 99\% pure) as precursors. Weighed powders were grinded in an agate mortar along with propan-2-ol to ensure homogenous mixing. The mixture was dried and pelletized under 4 tons hydraulic pressure. ZnGa$_2$O$_4$ (noted ZGO) and MgGa$_2$O$_4$ (noted MGO) pellets were annealed in air at 1300$^\circ$C and ZnAl$_2$O$_4$ (noted ZAO) pellet was annealed at 1400$^\circ$C for 6 hours, and later crushed to get fine powders for further characterizations. All three compounds were prepared with 1\% divalent cation deficiency and 0.5 mol\% Cr doping relative to trivalent cation. 

A Rigaku X-ray diffractometer was used to obtain the X-ray diffraction (XRD) patterns at room temperature (RT) using Cu-K$\alpha$ radiation. The spectra were recorded in 2$\theta$ range 20$^\circ$-80$^\circ$ with 0.02$^\circ$ step and 2$^\circ$/min scan speed. Rietveld refinement on the XRD patterns was carried out using FullProf software.\cite{35}

XAFS measurements in fluorescence mode have been carried out on Cr$^{3+}$ doped ZnGa$_2$O$_4$, MgGa$_2$O$_4$ and ZnAl$_2$O$_4$ compounds at Cr K edges. XAFS at RT were measured on the samples in fluorescence mode for Cr K edge at SAMBA beamline in Soleil synchrotron facility, France. Si (111) crystal plane was used as the monochromator. Absorbers were prepared by mixing 50 mg compound with 100 mg boron nitride and pressing each of them into 10 mm pellets. Photon yield was collected via Canberra 35 pixels SSD detector. XANES spectra were calculated using FEFF 8.4 software based on the self-consistent real-space multiple-scattering formalism.\cite{36} These calculations take into account multiple scattering of photoelectron from neighbouring atoms in presence of a fully relaxed core hole. Atomic coordinates were generated for the respective ideal spinel structure with their Rietveld refined lattice parameters, using ATOMS 2.5 version to obtain a FEFF output file.\cite{37} The respective octahedral core atom (Ga or Al) in the FEFF file was changed to Cr before running the FEFF program, to calculate the XANES pattern over a radius of about 8 \AA. The defects were then introduced accordingly around Cr in respective coordnates discussed later in the paper, and the spectra were computed for each case. These calculated spectra were linearly combined to obtain a best fit for the experimental spectra.

NIRPL decay were measured at RT on 180 mg powder samples filled into a 1cm sample holder. NIRPL emission was collected with a Scientific Pixis 100 CCD camera via an optical fiber linked to an Acton Spectra Pro 2150i spectrometer. The samples were excited at 560nm in the 4A2 - 4T2 absorption band of Cr$^{3+}$ by an optical parametric oscillator (OPO) EKSPLANT342B.

\section{Results and discussion}

\begin{figure}[h]
\centering
\includegraphics[width=\columnwidth]{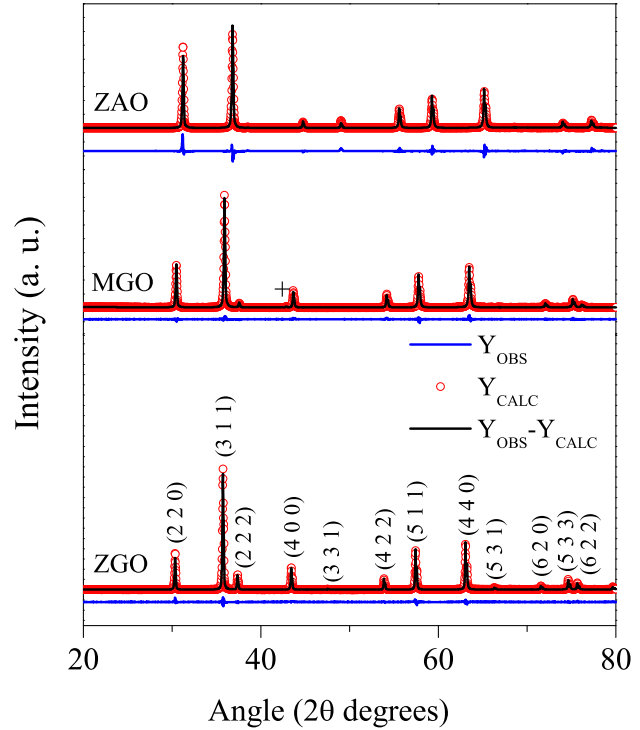}
\caption{X-ray diffraction patterns of Cr$^{3+}$ doped ZGO, MGO and ZAO compounds along with Rietveld refined fit and residual patterns. + in MGO pattern indicates small ($\sim$1\%) MgO impurity. }
\label{fig2}
\end{figure}

X-ray diffraction patterns indicated the formation of phase pure spinel compounds except for $\sim$ 1\% MgO impurity in MGO compound (Figure \ref{fig2}). Rietveld refinement was done on the XRD patterns with A site cations occupying tetrahedral 8a positions and B site cations occupying octahedral 16d positions.\cite{38} Lattice constant, cationic site occupancy along with scaling factor, background and instrumental parameters were adjusted to fit the experimental spectra. The resultant fit along with residual pattern for all compounds are shown in Figure \ref{fig2}. The cationic occupancy was not varied for ZGO compound while the Rietveld refinement yielded $\sim$45\% cationic site inversion in MGO in agreement with earlier reports.\cite{27,33} Varying site occupancy for ZAO compound yielded no cationic inversion hinting to the normal spinel lattice.\cite{17} The Rietveld refined lattice constants and the cationic occupancy are tabulated in Table \ref{table1}.

EXAFS results were presented in our earlier paper and were correlated with optical data.\cite{17} The bond distances of Cr to first neighbour (O in all cases), Cr to first cationic neighbour (Ga for ZGO and MGO, Al for ZAO) and Cr to second cationic neighbour (Zn for ZGO and ZAO, Mg for MGO), obtained by EXAFS fitting are given in Table \ref{table1}. For the EXAFS fitting of MGO, two models based respectively on normal spinel and inverse spinel structures were used. The bond distances given in Table \ref{table1} for MGO are the proportionate addition of the bond distances obtained from the above two models. Similar, but slightly different bond distances were calculated from the lattice parameters obtained using Rietveld refinement of XRD patterns and are also noted down in Table \ref{table1}. A preliminary comparison shows that the first and second cationic bond distances obtained from EXAFS fitting are almost similar to the XRD bond distances in all compounds.\cite{17} However EXAFS Cr-O bond distances in ZGO and MGO are much smaller compared to their respective XRD Ga-O bond distances. We had proposed in our previous paper that, this reduction in bond lengths is a result of the increased $\pi$ interaction of Cr and O orbitals due to the presence of antisite defects around Cr$^{3+}$.\cite{17} On the contrary, ZAO compound presents similar XRD Al-O and EXAFS Cr-O bond distances indicating no visible deformation in Cr$^{3+}$ local environment.

\begin{table*}[htbp]
\caption{Lattice constants, site occupancies and bond distances obtained from the Rietveld refinement of the XRD patterns, and bond distances obtained from the EXAFS fitting for Cr$^{3+}$ doped AB$_2$O$_4$ spinels (A = Zn, Mg and B = Ga, Al).}
\label{table1}
\begin{center}
\begin{tabular}{l|c|c|c|c}
\hline\hline
\multicolumn{2}{c|}{Parameters} & ZGO & MGO & ZAO  \\ [0.5ex] \hline 
\multirow{6}{*}{Rietveld refinement fitting} & Lattice constant (\AA) & 8.33291(3) & 8.28012(5) & 8.09372(6) \\ [0.5ex] 
                  & \multirow{2}{*}{A site occupancy} & 0.04125 (Zn) & 0.03702(14) (Ga) & 0.04125$^*$ (Zn) \\ [0.5ex] 
                  &                   &  & 0.00423(14) (Mg) &  \\ [0.5ex] 
                  & \multirow{3}{*}{B site occupancy} & 0.08291 [Ga] & 0.03702(14) [Mg] & 0.08291$^*$ [Al] \\ [0.5ex] 
                  &                   & 0.00042 [Cr] & 0.04589(14) [Ga] & 0.00042 [Cr] \\ [0.5ex] 
                  &                   &  & 0.00042 [Cr] &  \\ [0.5ex] \hline 
\multirow{3}{*}{Bond distances from XRD (\AA)} & Ga/Al-FN & 2.062 & 2.049 & 2.003 \\ [0.5ex] 
                  & Ga/Al-1CN & 2.946 & 2.927 & 2.862 \\ [0.5ex] 
                  & Ga/Al-2CN & 3.455 & 3.433 & 3.356 \\ [0.5ex] \hline 
\multirow{3}{*}{Bond distances from EXAFS fitting (\AA)} & Cr-FN & 1.96 (2) & 1.94 (2) & 1.992 (4) \\ [0.5ex] 
                  & Cr-1CN & 2.95 (3) & 2.91 (2) & 2.882 (7) \\ [0.5ex] 
                  & Cr-2CN & 3.42 (5) & 3.44 (2) & 3.350 (6) \\  [0.5ex]
\hline\hline
\end{tabular}
\end{center}
$^\ast$ d-ZAO pattern was Rietveld refined with only normal spinel structure model since an attempt to fit also the inverse spinel model yielded unphysical occupancies. \\
FN refers to first neighbour, 1CN refers to first cationic neighbour and 2CN refers to second cationic neighbour.
\end{table*}

\begin{figure}[h]
\centering
\includegraphics[width=\columnwidth]{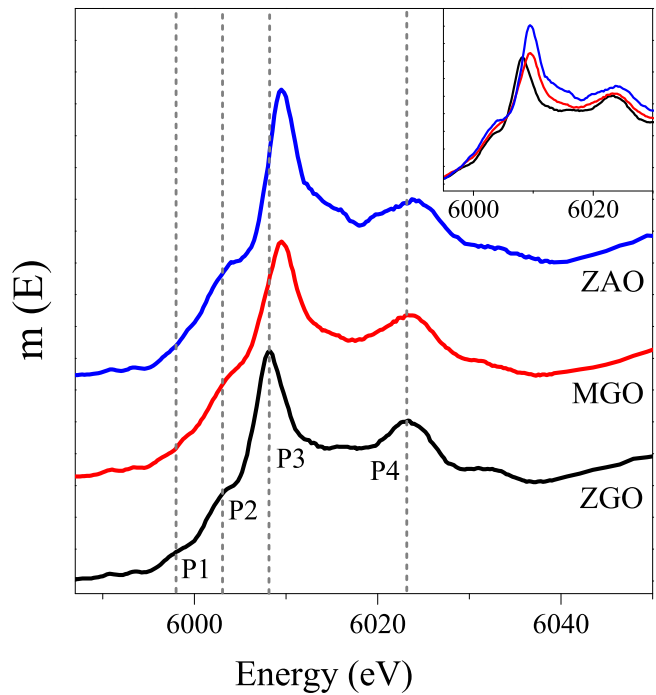}
\caption{Experimental Cr-edge XANES spectra for ZGO, MGO and ZAO compounds. Inset shows a zoom of the overlapping spectra with actual intensities.}
\label{fig3}
\end{figure}

The Cr K edge XANES spectra recorded for all the three discussed compounds are presented in Figure \ref{fig3}. Each spectrum shows four prominent features, P1 to P4 (marked by dotted lines for ZGO) in the near edge region. The features P1 and P2 arise due to transition of the photoelectron (e$_-$) from 1s to 4p orbitals of Cr, P3 is the main resonance peak arising due to the transition of e$_-$ from 1s to continuum and P4 represents the first constructive interference peak arising out of scattering of the e$_-$ with Cr neighbours. A comparison of the spectra shows a narrower and more intense P3 peak for ZAO as compared to ZGO and MGO compounds (see inset). As discussed in our previous paper, presence of an antisite defect in the neighbourhood of Cr$^{3+}$ decreases the site symmetry thereby resulting in a greater mixing of t2g and eg orbitals of Cr$^{3+}$.\cite{17} A greater admixture of the Cr 3d states with ligand orbitals decreases the transition probability due to lower density of unoccupied states. This observation is in agreement with our previous hypothesis about the presence of large number of antisite defects around Cr$^{3+}$ in ZGO and MGO as compared to ZAO.

In order to further investigate the presence of antisite defects around Cr ions, an attempt was made to calculate Cr K edge XANES spectra using FEFF 8.4 software. For these calculations, structural information obtained from Rietveld refinement was used as a starting point. Contribution of antisite defects to the Cr XANES spectra was estimated by using four different near neighbour environments for Cr ion. For example in ZnGa$_2$O$_4$, XANES spectra were calculated for Cr with: case(a) normal spinel environment (Ga at octahedral sites and Zn at tetrahedral sites) and Cr-O, Cr-1CN and Cr-2CN distances as obtained from Rietveld refinement, where 1CN and 2CN refer to first and second cationic neighbours; case(b) inverse spinel environment (Ga at tetrahedral site and Zn at octahedral site) and Cr-O, Cr-1CN and Cr-2CN distances as obtained from Rietveld refinement; case(c) inverse spinel environment with Cr-O, Cr-1CN and Cr-2CN distances as obtained from EXAFS, and case(d) normal spinel environment with Cr as first cationic neighbour instead of Ga. Case (d) corresponds to the presence of Cr$^{3+}$ clusters in neighbouring octahedral sites.

\begin{figure}[h]
\centering
\includegraphics[width=\columnwidth]{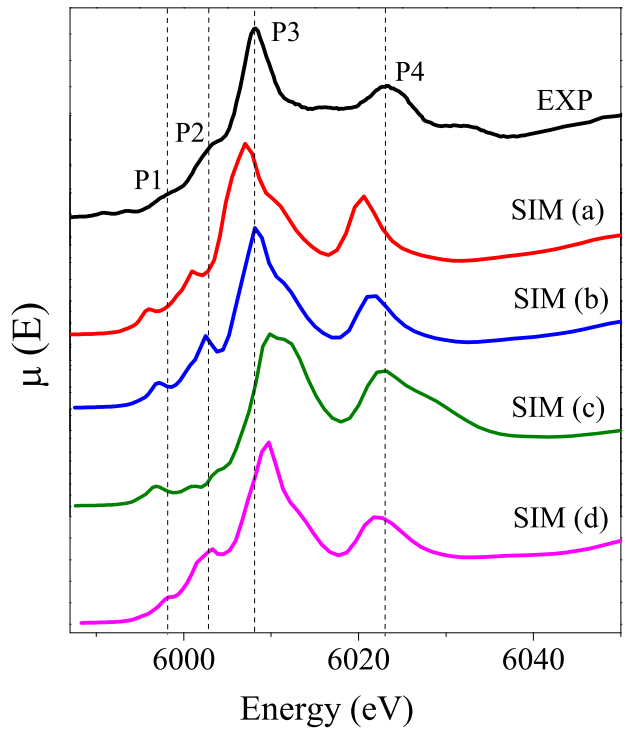}
\caption{Experimental Cr-K edge XANES spectrum (EXP) compared with spectra calculated with different environment for Cr- SIM (a) corresponds to the ideal ZGO structure with Cr core (Cr-O distance $\approx$ 2 \AA); SIM (b) corresponds to Cr with 6 neighbouring antisite defects (Cr-O distance $\approx$ 2 \AA); SIM (c) corresponds to Cr with 6 neighbouring antisite defects (Cr-O distance = 1.9  \AA); SIM (d) corresponds to Cr$^{3+}$ cluster (Cr-O distance $\approx$ 2 \AA).}
\label{fig4}
\end{figure}

All the calculated Cr-edge XANES spectra of ZGO are presented in Figure \ref{fig4} (see SIM (a) to (d)) and can be compared with the experimental spectrum (EXP). The curve SIM (a) in Figure \ref{fig4} was calculated with Cr in normal ZnGa$_2$O$_4$ lattice. It can be seen that all the four major features represented by P1 to P4 are present in this spectrum but their positions are shifted to lower energy. Furthermore the energy difference between any two of the features also does not match with that in experimental spectrum. It is important to note that such a difference between experimental and calculated spectrum hints at Cr$^{3+}$ environment being far from that expected for an ideal octahedron around Cr$^{3+}$ ions.

\begin{figure}[h]
\centering
\includegraphics[width=\columnwidth]{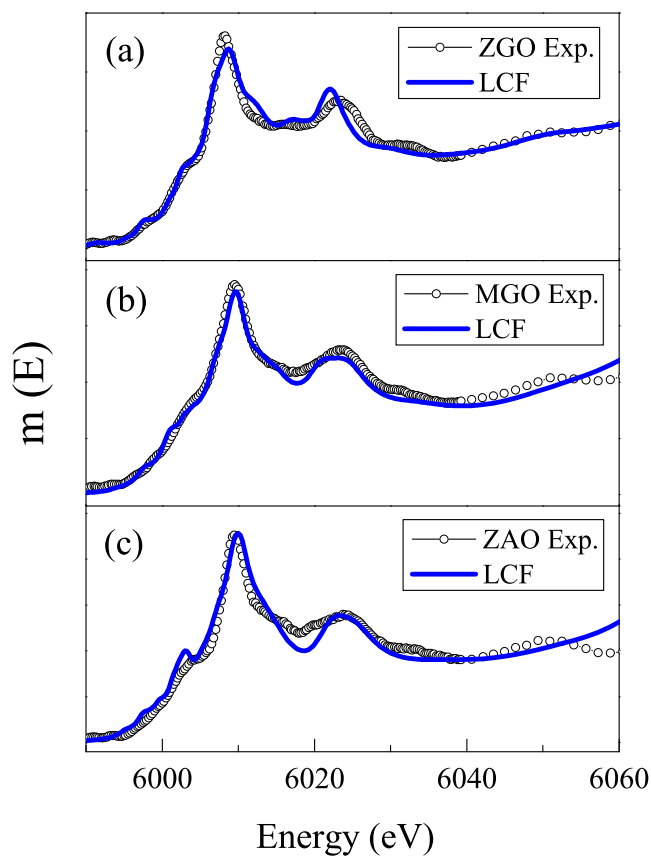}
\caption{Experimental Cr-K edge XANES spectrum along with computed LCF for (a) ZGO, (b) MGO, (c) ZAO.}
\label{fig5}
\end{figure}

Although the major features are also reproduced in Cr K edge XANES with inverse spinel environment (SIM (b)), this spectrum too cannot on its own account for the observed experimental spectrum. The major shortcoming being that the energy difference between P3 and P4 in this spectrum is quite less than that in the experimental spectrum. The third spectrum shown as SIM (c) in Figure \ref{fig4} is calculated using the nearest neighbour distances obtained from EXAFS fitting. Here, though the position of P4 matches with that in the experimental spectrum, P3 and P2 do not match. The fourth calculated spectrum (SIM (d) in Figure \ref{fig4}) represents the case of Cr clustering. This case was taken into account and was specially calculated as our EPR analysis performed on Cr doped MgGa$_2$O$_4$ and ZnGa$_2$O$_4$ clearly showed the presence of Cr clusters.\cite{8,17} In this case, though the positions of features P1, P2 and P4 are in fair agreement with those in experimental curve, the main resonance peak (P3) is shifted to higher energy in the calculated spectrum. A comparison of all the four calculated spectra with the experimental curve gives an indication that some or all of these curves could add up in some proportion to simulate the experimental curve. Studies on ZnGa$_2$O$_4$:Cr$^{3+}$ using optical spectroscopy, thermally stimulated luminescence and EPR spectroscopy have also shown that the Cr ions possess regular and distorted octahedral cages around them, as well as some Cr$^{3+}$ clusters.\cite{8} In order to estimate the contribution of each of the calculated spectra to the experimental curve, a linear combination fitting (LCF) was performed of all possible combinations of above calculated spectra using Athena software for XAFS analysis. The best fitted (lowest R factor) curves in case of three spinels are presented in Figure \ref{fig5}.

Figure \ref{fig5} (a) presents the experimental Cr XANES spectrum in ZGO and the best fitted (R factor = 0.019) curve calculated by LCF analysis. This computed curve is composed of 71\% contribution from spectrum of Cr in normal ZnGa$_2$O$_4$ environment (SIM (a) in Figure \ref{fig4}), nearly 17\% contribution from Cr in inverse spinel environment with EXAFS Cr-O bond distances (SIM (c) in Figure \ref{fig4}) and the rest of 12\% contribution from Cr clusters (SIM (d) in Figure \ref{fig4}). It may be noted that in the fitting, curve SIM (d) presents itself with a large shift ($\sim$6 eV) in its edge energy. The LCF curve with contributions from SIM (a) and SIM (c) in a ratio 75:25 also provides a fairly good agreement with experimental curve (R factor = 0.037). This indicates that Cr environment in ZGO is mostly that of a normal spinel, with about 25\% distorted octahedral environment arising due to cationic site inversion and chromium clustering. All these results fairly agree with EPR spectroscopy of ZGO, which shows the contribution of Cr$^{3+}$ in normal octahedral sites (more than 60\%), with contribution of Cr$^{3+}$ with neighbouring antisite defects and of antiferromagnetically coupled Cr$^{3+}$ clusters.\cite{8}

In the case of MGO, the best fitting LCF curve (Figure \ref{fig5} (b)) is composed of only 9\% of normal spinel environment, about 22\% of inverted spinel environment and the remaining 69\% contribution from spectrum representing Cr clustering. The two points to be noted here are; (i) significant increase in the Cr clustering in MGO as compared to that in ZGO and (ii) only 22\% of inverted spinel environment around Cr in spite of 44\% cationic site inversion found in the crystal lattice. The presence of anundant Cr clusters in MGO was also clearly indicated from EPR spectroscopy wherein the experimental X band Cr$^{3+}$ EPR signal was best simulated with a 1:3 ratio of isolated Cr ions subjected to inverted spinel environment and antiferromagnetically coupled Cr clusters.\cite{17} While the optical spectroscopy indicates that visible light induced NIRPL is entirely localized around Cr$^{3+}$ ion with an antisite defect as first cationic neighbour, its low intensity in MGO as compared to that in ZGO could possibly be due to presence of Cr clusters. 

\begin{figure}[h]
\centering
\includegraphics[width=\columnwidth]{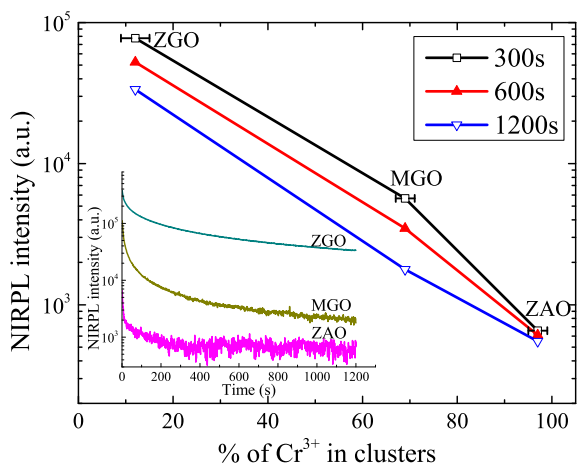}
\caption{Variation of NIRPL intensity values taken at different times versus the \% of Cr clusters present in ZGO, MGO and ZAO. Inset shows variation of NIRPL signal excited at 560nm as a function of time in Cr doped three spinel hosts.}
\label{fig6}
\end{figure}

The role of Cr clusters in decreasing visible light induced NIRPL is further strengthened by LCF analysis of Cr XANES in ZAO. This material exhibits very negligible NIRPL signal when subjected to visible light excitation although a quite strong NIRPL signal is observed with X-ray excitation.\cite{17} Here the best fitting LCF curve (Figure \ref{fig5} (c)) is made up of overwhelmingly large ($\sim$97\%) contribution from the spectrum calculated with Cr clusters (SIM(d) in Figure 4), while the rest 3\% is contributed by Cr in normal spinel environment with Al and Zn ions as first and second cationic neighbours respectively. A very good correlation exists between the \% of Cr clusters present around Cr$^{3+}$ ions in AB$_2$O$_4$ (A = Zn or Mg and B = Ga or Al) hosts and NIRPL intensity. In Figure \ref{fig6} values of NIRPL intensity at three different times in ZGO, MGO and ZAO are plotted against the respective \% of Cr in clusters. Strong similarities in correlations plotted at different times indicate the possibility that Cr clusters are responsible for complete quenching of the NIRPL excited by visible light. It should be stressed that although there was an important contribution of Cr$^{3+}$ clusters in XANES simulation, these clusters did not appear in the simulation of EPR spectrum of Cr$^{3+}$ in ZAO. The latter was fairly simulated with Cr$^{3+}$ ions in normal octahedral sites, without contributions of antisite defects and Cr clusters.\cite{17} However, it is possible to reconcile XANES and EPR results by recalling that these Cr$^{3+}$ clusters are weakly antiferromagnetically coupled in gallates (ZGO and MGO). The ground state of these clusters is characterized by a total electron spin S=0, so that they are EPR silent at very low temperature, while the EPR signal at high temperature is due to thermally accessible states with S$ > $0. Indeed the EPR signal of Cr clusters in ZGO and MGO decreases upon decreasing the temperature, and vanishes below 10 K.\cite{8} We may thus hypothesise that, antiferromagnetic coupling in ZAO is stronger than that in ZGO and MGO, so that these clusters remain EPR silent (S=0) even at room temperature, while they still dominate XANES spectrum.

\section{Conclusions}

The analysis of the Cr K edge XANES spectra presented here substantiates the previous EXAFS, EPR and optical results in establishing the presence of inversion defects around Cr$^{3+}$ ion and their role in persistent luminescence under low photon energy (visible light) charging process. Further the analysis not only indicates the defects around the trivalent chromium recombination centres are indeed associated with the low photon energy excitation persistent luminescence in AB$_2$O$_4$ type spinels, but also indicate the presence and the key role of Cr clusters. The main result of this work is that a clear correlation appears between the amount of Cr clustering and the quenching of the NIRPL emission intensity. In the aim of improving our understanding of NIRPL mechanism, it is thus important in the future to explore the consquences of this Cr-clustering by correlating this strong inhomogeneity of Cr$^{3+}$ distribution with thermally stimulated luminescence (TSL) and EPR spectra.

\section{Acknowledgment}

Authors acknowledge SAMBA beamline, Soleil synchrotron facility, for giving the beamtime. Ms. St\'{e}phanie Belin, beamline scientist, is gratefully thanked for the experimental support. Financial support from Indo-French Centre for the Promotion of Advanced Research (IFCPAR)/ CEntre Franco-Indien Pour la Recherche Avanc\'{e}e (CEFIPRA) is acknowledged.

\end{document}